\documentclass[aps,onecolumn,groupedaddress,superscriptaddress,longbibliography]{revtex4-2}
\usepackage{graphicx}  
\usepackage{dcolumn}   
\usepackage{bm}        
\usepackage{amsmath}
\usepackage{amssymb}   
\usepackage{xcolor}
\usepackage{subfigure}

\begin{document}
\title{Armstrong Liquid Bridge: Formation, Evolution and Breakup}

\author{Xueqin Pan}
\author{Man Hu}
\author{Bingrui Xu}
\author{Feng Wang}
\author{Peng Huo}
\author{Fangqi Chen}
\author{Zhibo Gu}
\author{Daosheng Deng}
\email{dsdeng@fudan.edu.cn}
\affiliation{Department of Aeronautics and Astronautics, Fudan University, Shanghai, 200433, China}
\date{\today}

\begin{abstract}
In this paper, we experimentally explore the formation, evolution and breakup of Armstrong liquid bridge. The extremely complicated evolution stage is revealed, which involves many coupled processes including the morphology change, current variation, heat transfer, and water evaporation. By focusing on the final fate of this liquid bridge, we observe that the breakup occurs once an effective length ($\tilde{L}$) is reached. This effective length increases linearly with the applied voltage, implying a threshold electric field to sustain the liquid bridge. Moreover, by an introduced external flow, the lifetime of the liquid bridge can be controlled, while the effective length associated with the breakup is independent on the external flow rate. Hence, these findings remarkably demonstrate that the breakup of liquid bridge is directly correlated with the effective length. In order to understand this correlation, a simplified model of an electrified jet is employed to take the electric field into account. By the linear stability analysis, the attained phase diagram agrees with the experiments well. Although a more comprehensive theory is required to consider other factors such as the surface charges, these results might provide a fresh perspective on the ``century-old" Armstrong liquid bridge to further elucidate its underlying physical mechanism.

\end{abstract}
\maketitle

\section{Introduction}
In 1893, Lord William G. Armstrong discovered a fascinating observation of ``a rope of water suspended between the lips of the two glasses'' under an applied voltage \cite{armstrongelectrical1893}. Recently, this Armstrong liquid bridge (ALB) has been revisited with the advanced experimental tools, such as the high-speed camera to reveal its rapid formation dynamics \cite{Fuchsfloating2007,Handwater2007,FuchsDynamics2008}. This system is extremely complicated, involving the coupled physical processes (EHD, heat transfer, and mass transport \cite{Morawetztheory2012}) and the undesirable chemical reactions (water splitting at the electrodes and the production of protons \cite{FuchsRaman2019}, and the change of pH \cite{WoisetschlExperiments2010}).

One of the distinctive features is that Armstrong liquid bridge floats horizontally or hangs stably between the two beakers without much deflection vertically in a gravity-defying fashion. This feature has been extensively studied using a simple classical electro-hydrodynamics (EHD) theory \cite{MelcherEHD1969,SavilleEHD1997}. Both the surface tension and Maxwell stress can provide the upward forces to counteract against the downward gravitational force, in order to hold ALB \cite{WidomTheory2009,Aerovwhy2011,Naminexperimental2013}.


\begin{figure*}[t]
	\centering
	\includegraphics[width=\textwidth]{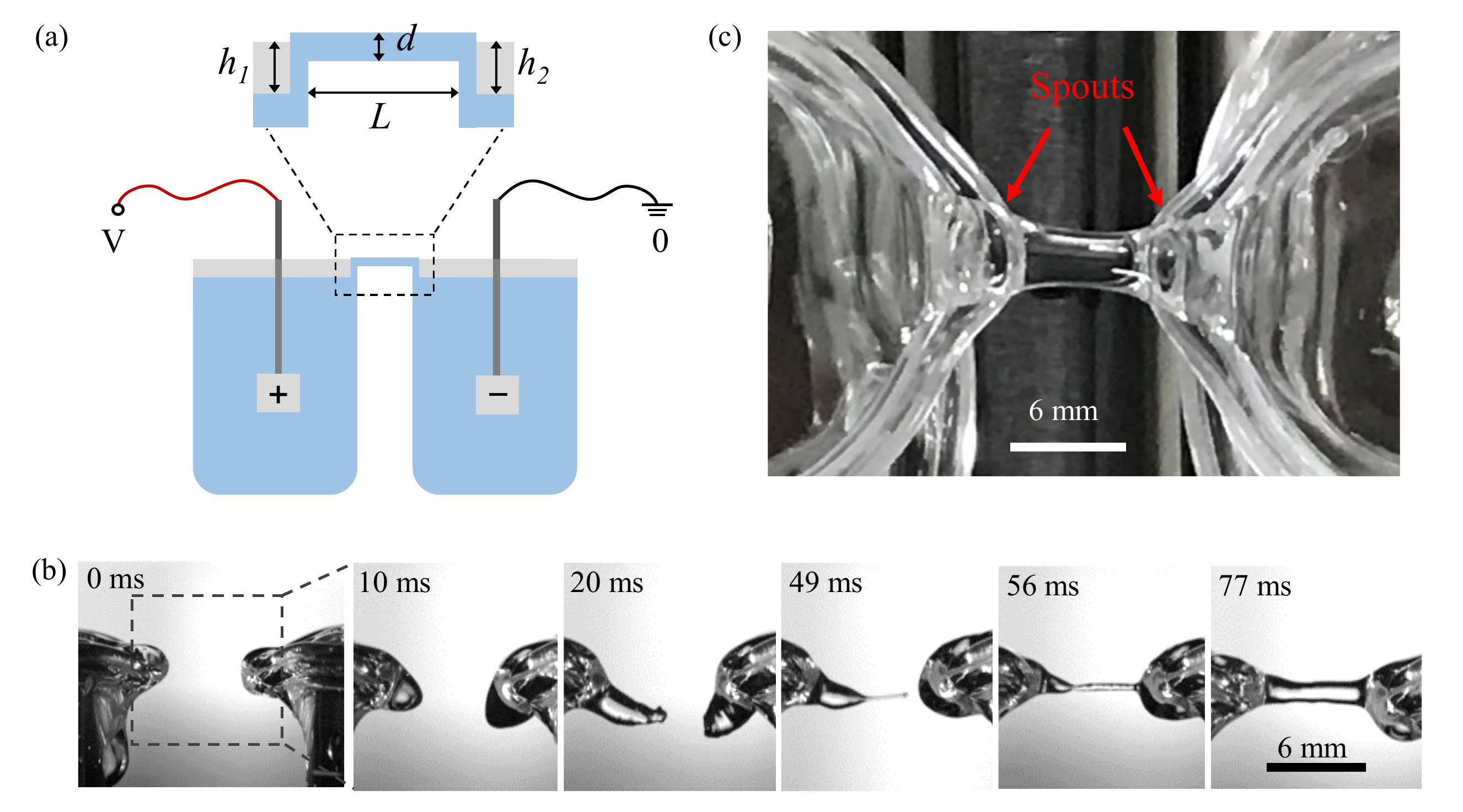}
	\caption{Formation of ALB. (a) The sketch of the experimental setup and a magnified view of ALB suspended between beakers with their small spouts facing closely ($V$ for the applied voltage, $L$ for the distance between the two spouts, $d=2r_0$ for the diameter, $h_1$ and $h_2$ for the climbing height, and the cylindrical beakers with 3.2-cm diameter and 5-cm height); (b) High-speed images for the initial formation of ALB from a side view ($L$ = 6 mm, $V$ = 13 kV); (c) a top view of the liquid bridge. }
\label{fig:formation}
\end{figure*}

Another intriguing feature is that Armstrong liquid bridge can persist or be stable for a while up to tens of minutes \cite{FuchsDynamics2008}. However, the final fate of Armstrong liquid bridge (the breakup and the associated lifetime), which is essential to elucidate the underlying physical mechanism for its stability or persistence, has not been thoroughly investigated at a quantitative level in the literature, to the best of our knowledge.

In this work, we focus on the stability of the floating bridge from the perspective of its final fate in terms of its lifetime once the breakup appears, and highlight three main findings. First, the breakup occurs once an effective length ($\tilde{L}$) is reached under a given voltage. Remarkably, this effective length has a simple linear relationship with the applied voltage. Secondly, by introducing an external flow, the lifetime of liquid bridge can be controlled, while the effective length associated with breakup is independent on the flow rate. Thirdly, in order to understand the observed correlation between an effective length and breakup, a theoretical model is employed and the linear stability analysis agrees with experiments well.

The paper is organized as below. In Section \ref{sec:formation}, the initial formation of ALB arises from the emission and connection of Taylor-cone jets. In Section \ref{sect:evolution}, the extremely complicated evolution stage is revealed, which involves many coupled processes, such as the morphology change, current variation, heat transfer, and water evaporation. Then in Section \ref{sect:breakup}, the occurrence of the breakup is found to be correlated with an effective length ($\tilde{L}$). In Section \ref{sect:controlflow}, an external flow is introduced to manipulate the liquid bridge and its lifetime ($\tau_\mathrm{lifetime}$), but the effective length associated with the breakup is unaffected by the external flow. In Section \ref{sect:theory}, a model of an electrified jet is proposed by taking the electric field into account, and a phase diagram is obtained to compare with the experiments. Lastly, the discussions and outlook for the ALB are provided in Section \ref{sect:discussions}.

\section{Formation of ALB: Emission and connection of Taylor-cone jets}
\label{sec:formation}
As shown in the sketch of Fig.~\ref{fig:formation}a, two cylindrical beakers (with the diameter and height about 3.2 cm and 5 cm, respectively) are filled with the deionized water (Shanghai Diena Biotechnology Co., LTD), and separated with a distance \textit{L} about millimeters while their small spouts facing each other closely in the opposite direction. An external voltage ($V$) through an high voltage power supplier (DW-P303-5ACF 1) is applied between the beakers through platinum electrodes immersed into water. As the voltage is gradually increased to a critical voltage ($V_c$) on the order of kV, the initial formation of ALB is revealed by the high-speed images (Phantom V611) [Fig.~\ref{fig:formation}b for a side view, and Supplementary Video (SV) 1]. Because of the gradual charge accumulation in water near the spouts under the applied voltage, water climbs or creeps to the spouts of the beakers, generating the inception of Taylor cones (t = 10 ms) \cite{TaylorDisintegration1964}. Then around 20 ms, the Taylor cones are subjected to the strong electric field, resulting in their deformation or extension. Around 49 ms, a fine jet is emitted horizontally away from one of the Taylor cones, leading to a cone-jet structure \cite{CloupeauElectrostatic1989}. Once the jet from the left beaker connects to the right beaker around 56 ms, a liquid bridge floats between the two beakers, while its diameter ($d = 2r_0$ measured at the center of the liquid bridge) increases shortly. Eventually around 77 ms, the liquid bridge has been fully developed and reaches a stable state with the diameter around millimeters, as shown in Fig.~\ref{fig:formation}c (a top-down view from a digital camera, Nikon D7200 with the lens of AF-S DX NIKKOR 16-85 mm). The linear increase of $V_c$ with $L$ indicates that a critical electric field is required to initiate the formation of ALB ($E_\mathrm{formation} \approx 1$ kV/mm)  (Appendix A).

\section{Evolution of ALB}
\label{sect:evolution}
In contrast to the extremely fast process within tens of milliseconds during the initial formation \cite{Fuchsfloating2007,FuchsDynamics2008,MariBuilding2010}, ALB can persist or be sustained for a much long period, up to tens of minutes. {This evolution process is extremely complicated and systematically investigated here, including the current, heat transfer, mass transfer, mass loss, and climbing heights.}
\begin{figure*}[t]
	\centering
	\includegraphics[width=\textwidth]{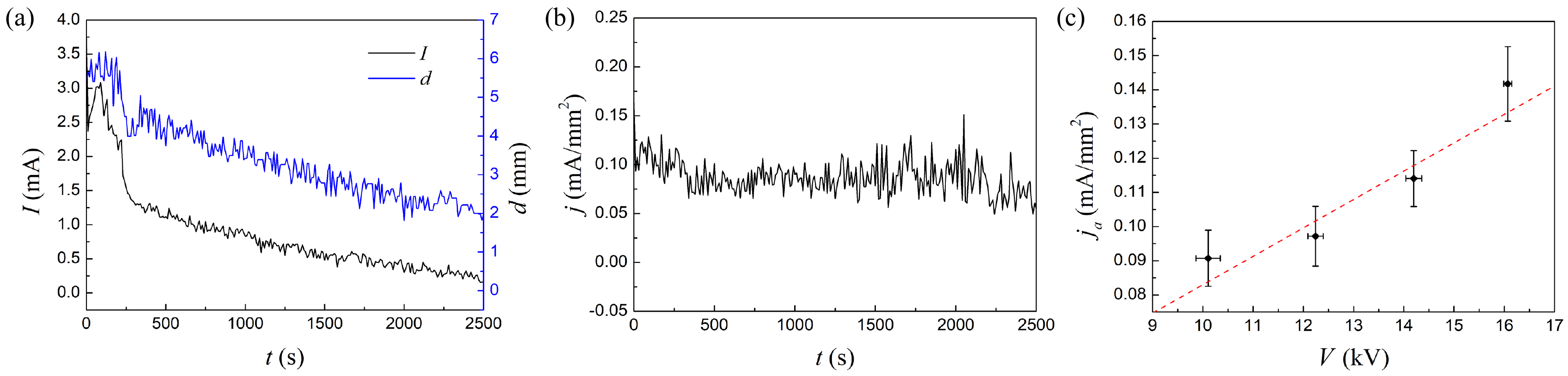}%
	\caption{{Current density. (a) The simultaneous measurement of current and diameter ($V=$ 12 kV, $L=$ 6 mm); (b) Current density $j$ fluctuating over the time; (c) The measurement of  $j_a$ (the black dots) agreeing with the calculation from the Ohm's law (the red dashed line). Error bars from four measurements.}}
\label{fig:current}
\end{figure*}

{\subsection{Current and diameter}}
{Once the liquid bridge is initiated, the water in the two beakers is connected by the liquid bridge to form an electric circuit under the applied voltage, generating the electric current. Current ($I$) changes with time, while the morphology and the associated diameter of liquid bridge evolves with time as well (Fig.~\ref{fig:current}a for $V=$ 12 kV). The current and diameter have a certain correlation: the smaller (larger) diameter has the higher (lower) resistance of the liquid bridge and the smaller (larger) current. Indeed, in the experiments, the current density ($j=I/A$) fluctuates over the time around the time-averaged current density ($j_a\approx 0.1$ mA/mm$^2$), as shown in Fig.~\ref{fig:current}b.}

{Additionally, $j_a$ has a tendency to increase with the applied voltage (Fig.~\ref{fig:current}c). From the Ohm's law,
\begin{equation}
j_a=\sigma E \propto  V,
\end{equation}
where $\sigma=0.5$  $\mu$S/cm is the conductivity of deionized water.  Then the calculated $j_a$ (the red dashed line) fits well with experimental data (Fig.~\ref{fig:current}c), indicating that the voltage is mainly dropped in the liquid bridge.  Since the resistance (proportional to the ratio between the length of the liquid bridge and the area of its cross section) of the liquid bridge is nearly 3 orders of magnitude higher than that of water in the beakers by estimating the geometric length scale, the voltage is reasonably expected to be dropped along the liquid bridge.}
\begin{figure*}[t]
	\centering
	\includegraphics[width=\textwidth]{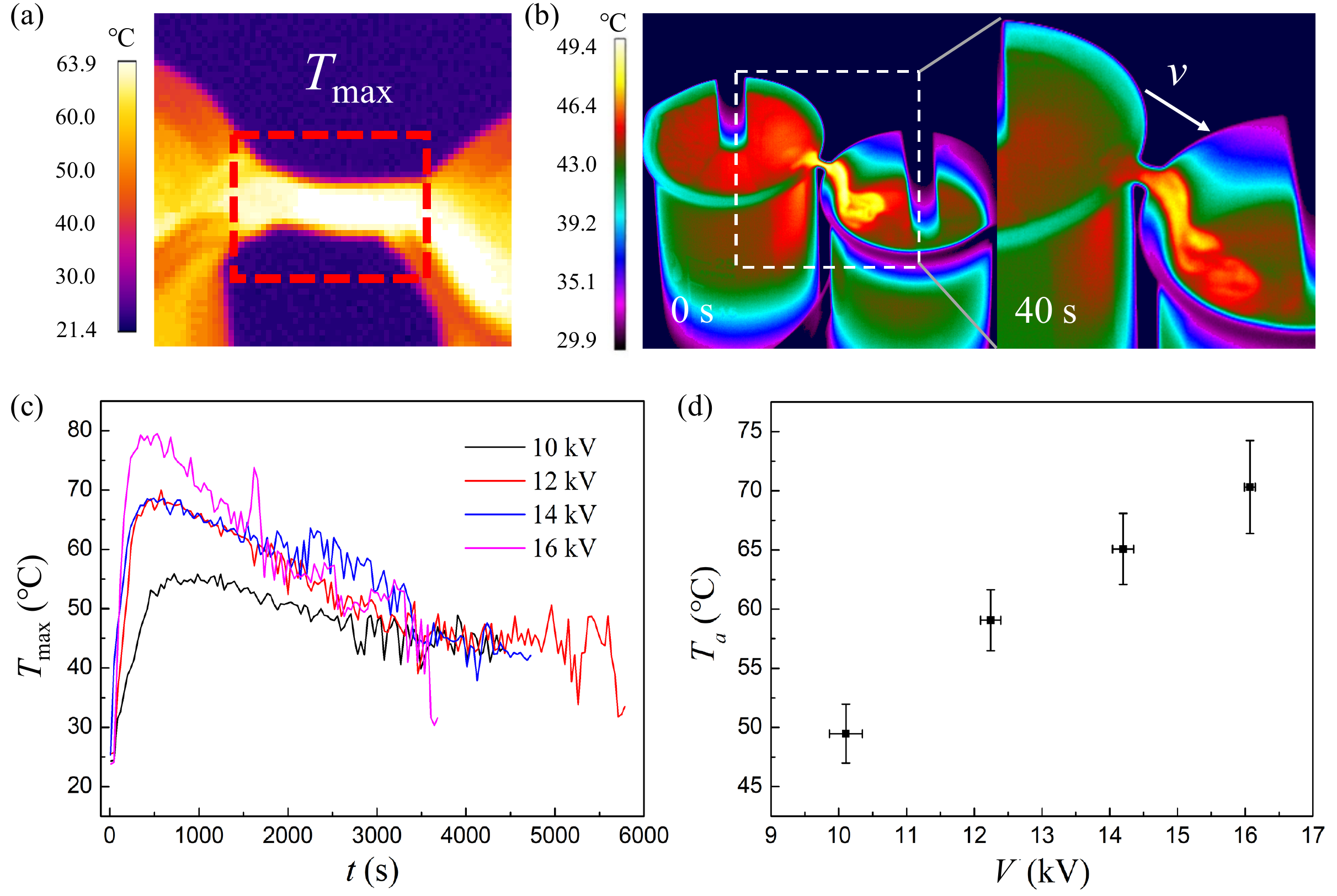}%
	\caption{{Temperature evolution. (a) Thermal image of the ALB ($V$ = 12 kV, $L$ = 6 mm); (b)thermal images revealing a waterfall falling from the liquid bridge to one of the beakers and the corresponding temperature distribution; (c) $T_{max}$ changing over time under different voltages; (d) $T_a$ increasing with the applied voltages (error bars for four measurements).}}
\label{fig:temeprature}
\end{figure*}

{\subsection{Heat transfer and temperature evolution}}
{Since the two beakers are connected by ALB, the water temperature (Fig.~\ref{fig:temeprature}a and b) is elevated due to the Joule heating, and is revealed by thermal images from the thermal camera (FLIR A6752sc). The temperature of liquid bridge is characterized by its maximum value along the liquid bridge ($T_{max}$), as indicated by Fig.~\ref{fig:temeprature}a. The evolution of $T_{max}$ and its dependence on $V$ are shown in Fig.~\ref{fig:temeprature}c. $T_{max}$ reaches its peak within ten minutes after connection and then begins to decrease gradually due to the effect of evaporation and heat dissipation. The time-averaged temperature ($T_a$) increases with the applied voltage(Fig.~\ref{fig:temeprature}d).}

\subsection{The reduction of water level and the increase of the climbing height}
\begin{figure*}[t]
	\centering
	\includegraphics[width=\textwidth]{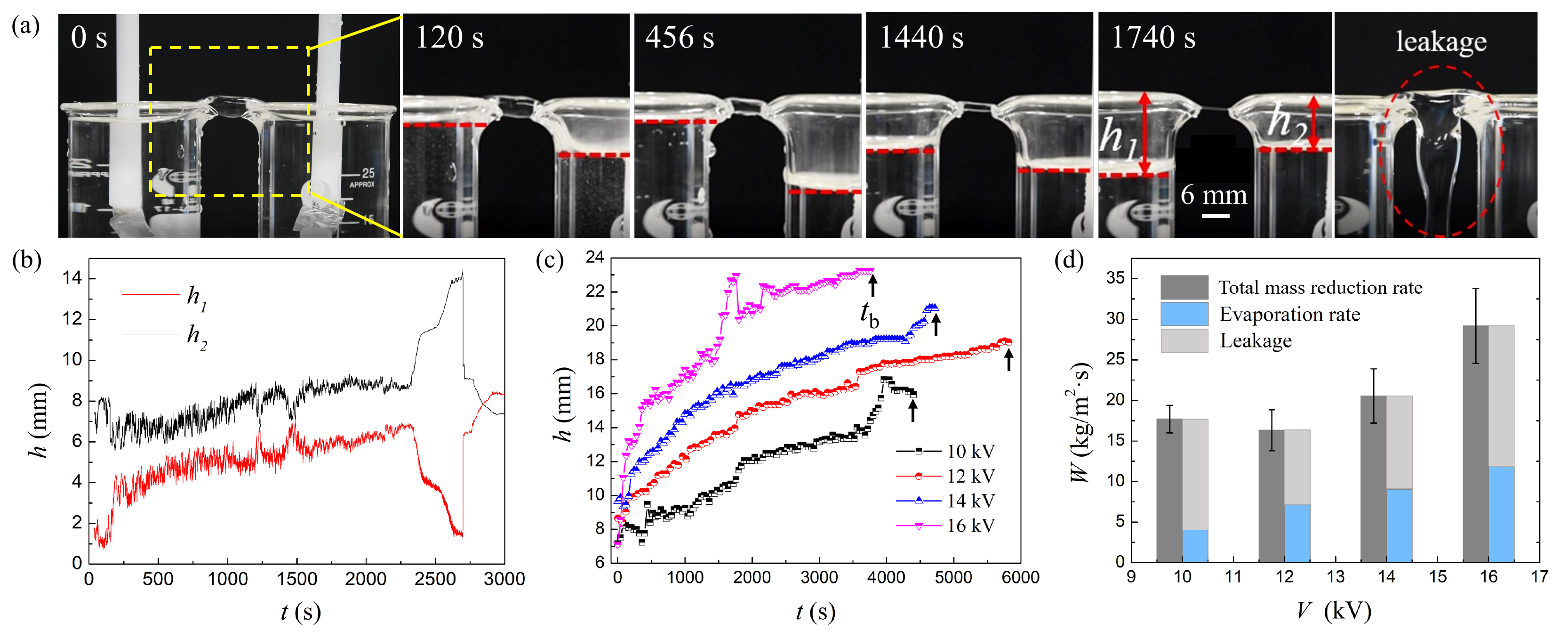}%
	\caption{Mass loss characterized by the reduced water level or the increased climbing height. (a) Photographs showing the change of water level (red dashed lines) inside the beakers with time, and the inadvertent leakage ($L$ = 6 mm, $V$ = 16 kV); (b) the climbing height of water level in two beakers at the anode side ($h_1$) and the cathode side ($h_2$) changing with time; (c) the total climbing height ($h=h_1+h_2$) increasing with time for various voltages; (d) the mass reduction of water due to the evaporation and the leakage (error bars for four measurements).}
\label{fig:evolution}
\end{figure*}

During the evolution stage, the surface level of water (red lines) in the beakers is gradually diminished about several millimeters (Fig.~\ref{fig:evolution}a and SV 2). Fig.~\ref{fig:evolution}b presents the change of the climbing height (the distance between the rim or the upper edge of the beaker and the water level, referring to the creeping height in \cite{Morawetztheory2012}) at the anode ($h_1$) and cathode side ($h_2$) (as shown in Fig.~\ref{fig:formation}a and Fig.~\ref{fig:evolution}a). The variations of $h_1$ and $h_2$ (in Fig.~\ref{fig:formation}a) is caused by an axial flow of the liquid bridge between the two beakers, forming a water fall (the thermal video in SV 3 and  Fig.~\ref{fig:temeprature}b).  This axial flow can be bidirectional, flowing from the anode (cathode) side to the cathode (anode) side between the beakers during the evolution \cite{WoisetschlExperiments2010, Morawetzreversed2017}. The velocity of this axial flow is on the order of tens of millimeters per second (more details in Appendix B).

Then the reduction of water level is characterized by the total climbing height ($h= h_1 + h_2$). As presented in Fig.~\ref{fig:evolution}c, $h$ gradually increases with time for various applied voltages until the final breakup time($t_b$). This reduction of water level corresponds to the total mass loss of water in the beakers.

\subsection{Mass loss: evaporation and leakage}
As shown in Fig.~\ref{fig:evolution}d, this average reduction rate of total mass ($W$, the dark shadow) is attributed to the evaporation (the blue shadow) and leakage (the light shadow). The evaporation of water inside the beakers increases with the applied voltage due to Joule heating. Based on the measured temperature, the evaporation rate (Fig.~\ref{fig:evolution}d) is theoretically estimated from Dalton's evaporation law \cite{1993Evaporation},
\begin{equation}
E=A[\frac{P(T_w)}{T_w}-\frac{HP(T_0)}{T_0}],
  \label{equ:evaporationrate}
\end{equation}
where $E$ is the evaporation rate, $T_w$ is the temperature of the water surface, $T_0$ is the temperature of the air atmosphere, $P(T)$ is the vapor pressure at temperature $T$, $H$ is the relative humidity in the air. $A$ is a constant dependent on the air velocity and temperature in experiments (0.3 m/s and 22$^{\circ}$C). Here the time-averaged temperature ($T_a$) in Fig.~\ref{fig:temeprature}d is approximately used as $T_w$. The calculated  evaporation rate under the applied voltage is about $1/3$ of the averaged rate of the total mass change (Fig.~\ref{fig:evolution}d).

The leakage of water takes place inadvertently during the evolution of the liquid bridge, as shown by the right snapshot in Fig.~\ref{fig:evolution}a and the supplementary video 2. Once water falls from the beakers or the liquid bridge down to the ground due to the gravity, the total mass of water (in the breakers and within the liquid bridge) is reduced accordingly, resulting in the change of the total climbing height. The leakage of water during the complicated evolution process of ALB takes place inadvertently or accidentally, and it is difficult to be predicted theoretically or described accurately due to its uncertain occurrence. Based on the assumption that the mass loss is due to the evaporation and leakage, the average leakage rate is obtained by subtracting the evaporation rate from the average rate of mass reduction. As shown in Fig.~\ref{fig:evolution}d, the average rate of the leakage is about $2/3$ of the averaged rate of the total mass reduction.

~\\
\section{Breakup of ALB}
\begin{figure*}[t]
	\centering
	\includegraphics[width=\textwidth]{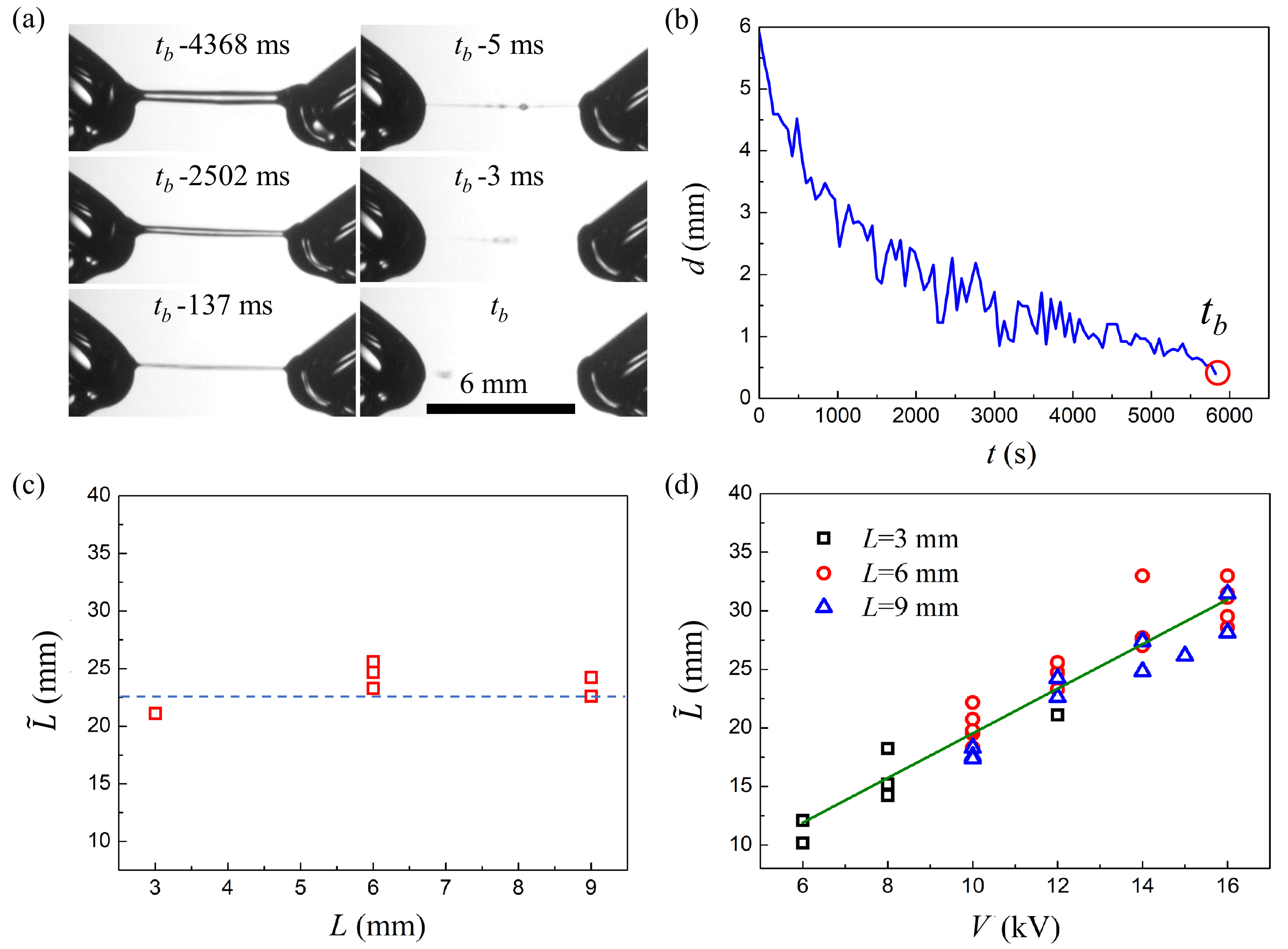}%
	\caption{Breakup of ALB. (a) High-speed images of the typical breakup ($L$ = 6 mm, $V$ = 12 kV); (b) diameter of bridge gradually decreasing until the final breakup; {(c) at the breakup time ($t_b$), $\tilde{L}=h(t_b)+L$ almost a constant for various $L$ ($V$ = 12 kV);} (d) experimentally $\tilde{L}$  increasing linearly with $V$ for various $L$, and its slope indicating a constant electric field $E_\mathrm{breakup}= 0.52$ kV/mm associated with the breakup.}
\label{fig:breakup}
\end{figure*}

Therefore, the gradual loss of water manifestly causes the continuous increase of $h$, and eventually results in the final fate of the breakup (Fig.~\ref{fig:breakup}a and b, and SV 4).
\label{sect:breakup}
{\subsection{Breakup dynamics}}
As captured by the high-speed images (Fig.~\ref{fig:breakup}a), the continuous liquid bridge within last several seconds is rapidly thinned to an extremely tiny filament to be more susceptible to the instability, eventually breaking up into a series of tiny droplets within the last five milliseconds.

During the whole course of ALB, the diameter of liquid bridge is decreased from six millimeters at the initial stage down to hundreds of micrometers approaching the breakup (Fig.~\ref{fig:breakup}b), as limited by the resolution of the high-speed camera experimentally. Although the radius near the spouts of the beakers at the initial stage depends on the electric field \cite{Morawetztheory2012,Naminexperimental2013}, this time-dependent diameter of the ALB  requires more future investigation.

{\subsection{An effective length ($\tilde{L}$)}}
{For a given voltage ($V=12$ kV), by carrying out a series of experiments, we find out at breakup time $t_b$ the total climbing height $h(t_b)$ decreases with $L$. Then an effective length is simply defined as below,
\begin{equation}
\tilde{L}=h(t_b)+L.
\end{equation}
As shown in Fig.~\ref{fig:breakup}c, this effective length ($\tilde{L}$) is nearly a constant or independent on the distance between beakers, $L$. In other words, at a shorter $L$, the breakup occurs at a larger $h(t_b)$; while at a longer $L$, the breakup occurs for a smaller $h(t_b)$. This reproductivity of the effective length is striking in the final fate of liquid bridge, considering such complicated evolution processes and uncertain factors such as the leakage.}

Moreover, we perform a series of experiments to investigate both $L$ and $V$ together, and identify $h(t_b)$ at the final breakup time. Noticeably, all the data of the effective length [$\tilde{L}=h(t_b)+L$] and $V$ (Fig.~\ref{fig:breakup}d) linearly collapse into a master line. Considering the complicated system of the liquid bridge under a high voltage (involving the heat transfer, flow, evaporation, the occasional leakage, and the chemical reactions), this correlation between an effective length and the voltage at the final fate of Armstrong liquid bridge is remarkable.


~\\

\section{Manipulation of ALB via an external flow}
\label{sect:controlflow}
In order to further validate the crucial role of the effective length at the final fate of ALB, an external flow is introduced to manipulate the change of water level and the associated lifetime.

\subsection{$\tilde{L}$ independent on the flow rate}
\begin{figure*}[t]
	\centering
	\includegraphics[width=\textwidth]{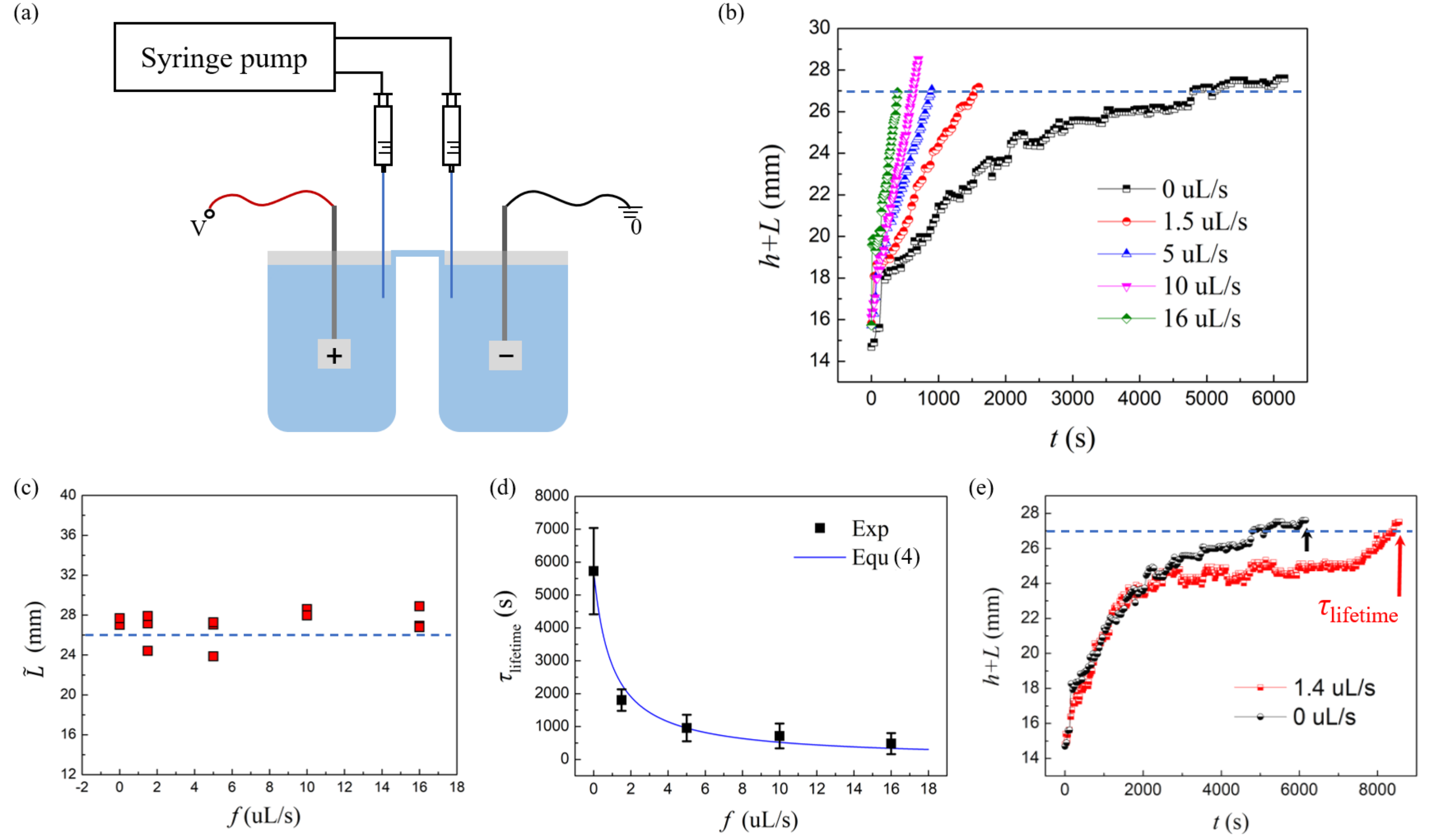}%
	\caption{Control of ALB by an external flow. (a) The sketch of manipulating liquid bridge by an external flow through a syringe pump; (b) the evolution of $h+L$ with time under various flow rates; (c) the similar $\tilde{L}=h(t_b)+L$  upon breakup, nearly independent on the flow rate; (d) the reduction of $\tau_\mathrm{lifetime}$ by the flow extraction (error bars for three measurements); (e) the extension of $\tau_\mathrm{lifetime}$ over two hours by flow injection ($V$ = 14 kV, $L$ = 6 mm).}
\label{fig:flowrate}
\end{figure*}

Water is extracted out of the beakers through syringe pumps (Ristron RSP04-BD) (Fig.~\ref{fig:flowrate}a). Each syringe (5 ml in volume) is pushed or pulled automatically at a given flow rate. One end of PTFE soft pipes is connected to the syringe, while the other end is immersed into water in beakers. By this way, the water can be extracted out of or injected into the beakers under various flow rates during the ALB experiments.

For a given voltage ($V=$14 kV), the time-dependent $h+L$ under different rates of extraction flow is presented in Fig.~\ref{fig:flowrate}b, and again the value of $h+L$ at the final breakup time $\tilde{L}=h(t_b)+L$  is nearly a constant (Fig.~\ref{fig:flowrate}c).

~\\
{\subsection{Reduced $\tau_\mathrm{lifetime}$ via flow extraction}}
The extraction flow correspondingly causes the increased $h+L$, resulting in the decreased $\tau_\mathrm{lifetime}$ (Fig.~\ref{fig:flowrate}d). Actually, this effect of external flow rate on $\tau_\mathrm{lifetime}$ can be estimated from the reduction of the mass by considering the evaporation rate ($C_e$), the flow rate of extraction ($f$), and the leakage mass ($\Delta C$). {For the case without the flow extraction, $\Delta m=\Delta C+ C_e\tau_0$, where $\Delta m$ is the total mass reduction of water during the whole experiment, $C_e$ is the evaporation rate, and $\tau_0$ is the lifetime without extraction. Then for the case with the flow extraction, $\Delta m=\Delta C+ C_e\tau+f\tau$, where $f$ is the extraction rate, and $\tau$ is the associated lifetime.}

{Despite the different extraction rates, when the beaker distance $L$  and the applied voltage $V$ are fixed, the total height or mass change is nearly a constant once the breakup occurs. Thus the total mass reduction $\Delta m$ is supposed to be the same in both cases. Under an assumption of a constant leakage $\Delta C$, the relation between $\tau$ and $f$ is obtained as below,}
\begin{equation}
\tau_\mathrm{lifetime}=C_e\tau_0/(C_e+f).
\label{equ:lifeflow}
\end{equation}
The above equation states that the lifetime $\tau$ is related with an evaporation rate $C_e$ and the applied extraction rate $f$. Indeed, the experimental data can be fitted well with Equation (\ref{equ:lifeflow}) (the blue line in Fig.~\ref{fig:flowrate}d). 

~\\
{\subsection{Extended $\tau_\mathrm{lifetime}$  via flow injection}}
{Conversely, by supplying water into beakers through injection to balance the water loss due to evaporation or leakage , the level of water surface is slowly changed, consequently taking a longer time to reach $\tilde{L}$ or extending the lifetime of liquid bridge. Fig.~\ref{fig:flowrate}e shows the evolution of $h+L$ without an external flow and with an injection flow rate at 1.4 $\mu$L/s. At beginning from 0 to 2,000 seconds, the injection has less effect on the climbing height. Between 2,000 s and 7,000 s, $h+L$ increases slowly due to the injection of water. However, after 7,000 seconds as the water in the syringe is used up, $h+L$ starts to increase fast due to the evaporation and leakage. Clearly, $\tau_{lifetime}$ has been extended up longer than 2 hours with an injection rate. For both cases, breakup occurs when $\tilde{L}$ reaches around 27 mm.}

~\\
\section{A theoretical model}
\label{sect:theory}
The final fate of Armstrong liquid bridge has been experimentally characterized by the effective length, $\tilde{L}$, which is independent on the beaker distance (Fig.~\ref{fig:breakup}c) and the external flow rate $f$ (Fig.~\ref{fig:flowrate}c). Remarkably, $\tilde{L}$ linearly increases with the voltage $V$ (Fig.~\ref{fig:breakup}d), and its slope implies an electric field $E_\mathrm{breakup} \approx 0.52$ kV/mm, below which the liquid bridge will experience the instability and be subjected to the breakup. In other words, a sufficient strong electric field might be necessary, if not sufficient, to maintain the liquid bridge or sustain its stability.

\begin{figure}[t]
	\centering
	\includegraphics[width=\textwidth]{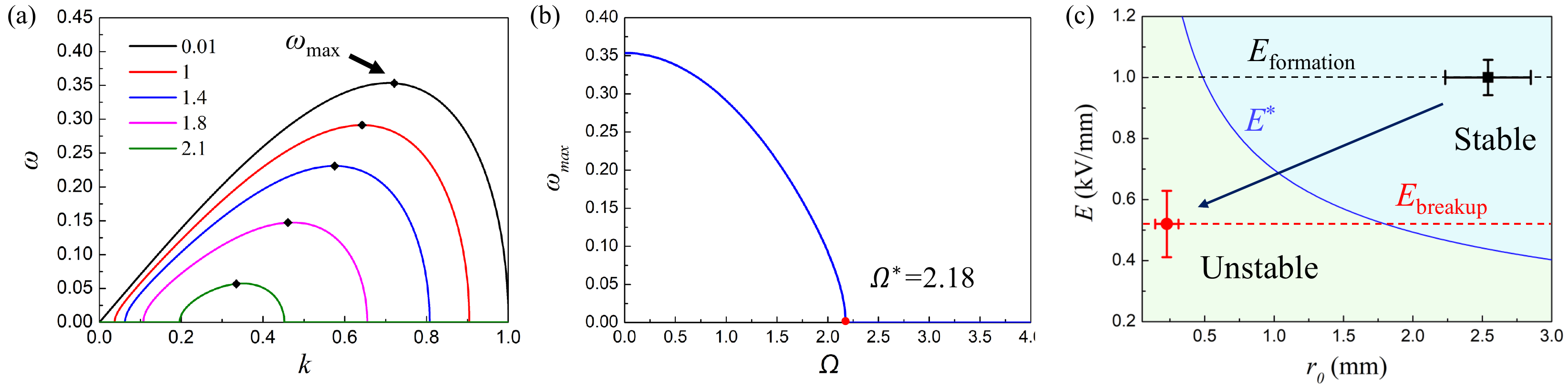}%
	\caption{Linear stability analysis. {(a)The dispersion relationship or the growth rate $\omega$ dependent on the wavelength $k$ under different dimensionless electrical field $\Omega$. $\Omega = 0.01, 1, 1.4, 1.8, 2.1$ from the outer line to the inner line, and $\omega_{max}$ for the maximum of $\omega$. }(b) The dispersion relationship ($\omega_{max}-\Omega$) showing a critical value of the dimensionless electric field ($\Omega^{\ast}=$2.18). (c) the calculated phase diagram ($E-r_0$) consistent with the experimental observation. The dark square for the initial formation, while the red circle for the final breakup (error bars from experimental measurements).}
\label{fig:theory}
\end{figure}

In order to understand the possible role of electric field for the stability, the complicated morphology of the liquid bridge in a quasi-static state \cite{Morawetztheory2012} is approximately simplified as an axi-symmetric or cylindrical liquid jet under an applied electric field. Then the model of an axial electrified jet \cite{SavilleEHD1970,Hohmanstability2001} is employed here, and a phase diagram is obtained by the linear stability analysis.

\subsection{{Dispersion relationship}}
The linear stability analysis allows the investigation of its stability on the perturbation, giving rise to the following dispersion relation between the perturbation growth rate ($\omega$) and the wavelength ($k$),
\begin{equation}
\omega=k \sqrt{\frac{1}{2}(1-k^2)-\frac{\Omega^2}{4 \pi} \left [ 1+\frac{2}{\beta k^2 \ln (Bk)}\right]},
\label{equ:maxgrowthrate}
\end{equation}
in the limit of the infinite conductivity, zero viscosity, and zero {surface} charge \cite{Hohmanstability2001}. Indeed this limit is applicable here, since the dimensionless conductivity is $K^{*}=K \sqrt{r^3\rho/(\gamma \beta)}/\bar{\epsilon} \sim 10^{4}$ and dimensionless viscosity $\mu^{*}=\sqrt{\mu/(\rho \gamma r)} \sim 10^{-3}$ ($\rho$ for the density of water, $r$ for the typical length scale, and $\mu$ for the kinetic viscosity of water). Here the dimensionless number $\beta=\epsilon/\bar{\epsilon}-1$, the growth rate is nondimensionalized with time scale $\sqrt{\rho r^3 /\gamma}$, whereas the dimensionless electrical field $\Omega=E/\sqrt{\gamma/[(\epsilon-\bar{\epsilon})r]}$, and $B=0.89$	for a constant.

{Based on Equation (\ref{equ:maxgrowthrate}), Fig.~\ref{fig:theory}a presents this dispersion relation between the perturbation growth rate ($\omega$) and the wavelength ($k$) under various dimensionless electric field $\Omega$. For each $\Omega$, there is a maximum growth rate of $\omega$ ($\omega_{max}$) for the wavelength in the range from 0 to 1, and $\omega_{max}$ decreases with $\Omega$.}

Then the maximum growth rate $\omega_{max}=max[\omega(k)]$ dependent on the electric field ($\Omega$) is presented in Fig.~\ref{fig:theory}b. $\omega_{max}$ gradually decreases with $\Omega$ due to the suppression of instability by the electric field, and eventually becomes zero associated with the marginal stability at $\Omega^{\ast}=$2.18, indicating the transition from instability to stability.

\subsection{Phase diagram}
Then from the dimensionless $\Omega^{\ast}$, the phase diagram between the critical electric field $E^{\ast}$ and the radius $r_0$ (the blue line in Fig.~\ref{fig:theory}c) is obtained as below,
\begin{equation}
\Omega^{\ast}=E^{\ast}/\sqrt{\gamma/[(\epsilon-\bar{\epsilon})r_0]}.
\label{equ:expelectricfield}
\end{equation}

To check the theory with experiments, both the aforementioned electric field at the initial formation ($E_\mathrm{formation}$, the black line) and the final breakup ($E_\mathrm{breakup}$, the red line) are indicated in this phase diagram (Fig.~\ref{fig:theory}c). During the evolution of liquid bridge within tens of minutes or hours, the electric field gradually decreases, while the diameter is reduced as well, consequently leading the transition from the stable regime at the initial stage (the dark square) to the unstable regime subjected to the final breakup (the red circle). Hence, the experiments are consistent with the theory well.
~\\

\section{Discussions and Outlook}
\label{sect:discussions}
We have presented our experimental findings of the final fate of Armstrong liquid bridge. Generally, this intriguing ALB is extremely complicated, and several more issues are still required to be further explored in future. First, although the observed effective length ($\tilde{L}$) associated with breakup is simple, the dynamics of climbing height ($h$) is determined by many complicated processes. For example, the current density leads to the elevated temperature due to the Joule heating, and the subsequent temperature-dependent evaporation causes the change of $h$. Also, other parameters (such as the axial flow of the liquid bridge and the accident water leakage) can affect $h$ as well.

Secondly, during the evolution stage, the bi-directional axial flow was observed using magnetic resonance imaging \cite{Wexler2017flow}, and was explained by the electro-osmotic flow due to the charges on the outer surface of the liquid bridge \cite{Morawetzreversed2017}. Here the axial flow in the liquid bridge is directly visualized by the ``water fall" from the thermal images (Fig. \ref{fig:temeprature}b), and its bi-directionality is clearly shown by the thermal video (SV 3) and the mass transport in the beakers (more details in Appendix B). Specifically, the maximum velocity \cite{Wexler2017flow} is comparable with the maximum flow velocity in order of tens of millimeters per second (Fig. \ref{fig:masschange}b). Since the temperature gradient exists along the liquid bridge as seen from the thermal images, the thermocapillary Marangoni flow might be another possible mechanism for this axial flow.

Thirdly, the simplified model of an electrified jet in Section \ref{sect:theory} mainly demonstrates the prerequisite electric field to stabilize the liquid bridge, in order to understand the correlation between the observed effective length with the breakup. For a more quantitative understanding, other effects including the aforementioned axial flow, the surface charges, and the excess ions in water (water splitting at the high voltage $\sim$ kV \cite{WoisetschlExperiments2010,Fuchs2016excesscharge}) are needed to be carefully examined in the model, which is beyond the scope of this work.

Fourthly, from the microscopic perspective, the existence or survival of ALB has been mainly attributed to the peculiarity of water molecules \cite{FuchsCan2010,FuchsArmstrongrev2014,SunAdCol2020, Giudicecollectivewater2010,Giudiceinfluencewater2011}. But this possible mechanism has been ruled out experimentally, since the water molecules do not exhibit the preferred orientation along the strong electric field in ALB from the measurement of neutron scattering, Raman scattering, and X-ray scattering data \cite{FuchsNeutron2009,PonterioRaman2010,Skinnerstructure2002}.
Nevertheless, here we identify that the breakup is directly correlated with an effective length ($\tilde{L}$) proportional to the applied voltage, implying that a threshold electric field ($E_\mathrm{breakup}$) is necessary to sustain the liquid bridge or be responsible for its stability, although the microscopic mechanism requires more careful research in future.

Lastly, the lifetime of ALB might be controlled by many other factors. For example, the ambient environments (such as the temperature and humidity) affect the evaporation rate to influence the water loss and its lifetime. 

In summary, we experimentally explore the formation, evolution and breakup of Armstrong liquid bridge, and report three main findings. First, the breakup occurs once an effective length ($\tilde{L}$) is reached, which is well reproduced under a given voltage and linearly increases with the applied voltage. Secondly, by introducing an external flow, the lifetime of liquid bridge can be controlled, while the effective length associated with the breakup is independent on the external flow rate. Hence, these experiments remarkably demonstrate that the final fate of liquid bridge is directly correlated with the effective length. Thirdly, in order to understand the observed stability and lifetime associated with the effective length, a simplified model of an electrified jet is employed by taking the electric field into account, and the linear stability analysis agrees with experiments well. These results provide a fresh perspective on the ``century-old" Armstrong liquid bridge --- a remarkable observation of the effective length $\tilde{L}$ is directly correlated with the breakup, shedding light on its underlying physical mechanism.


\appendix

\section{Initial morphology under various $L$ and the associated $V_c$}
\label{app:initialmorph}
\begin{figure*}[b]
	\centering
	\includegraphics[width=\textwidth]{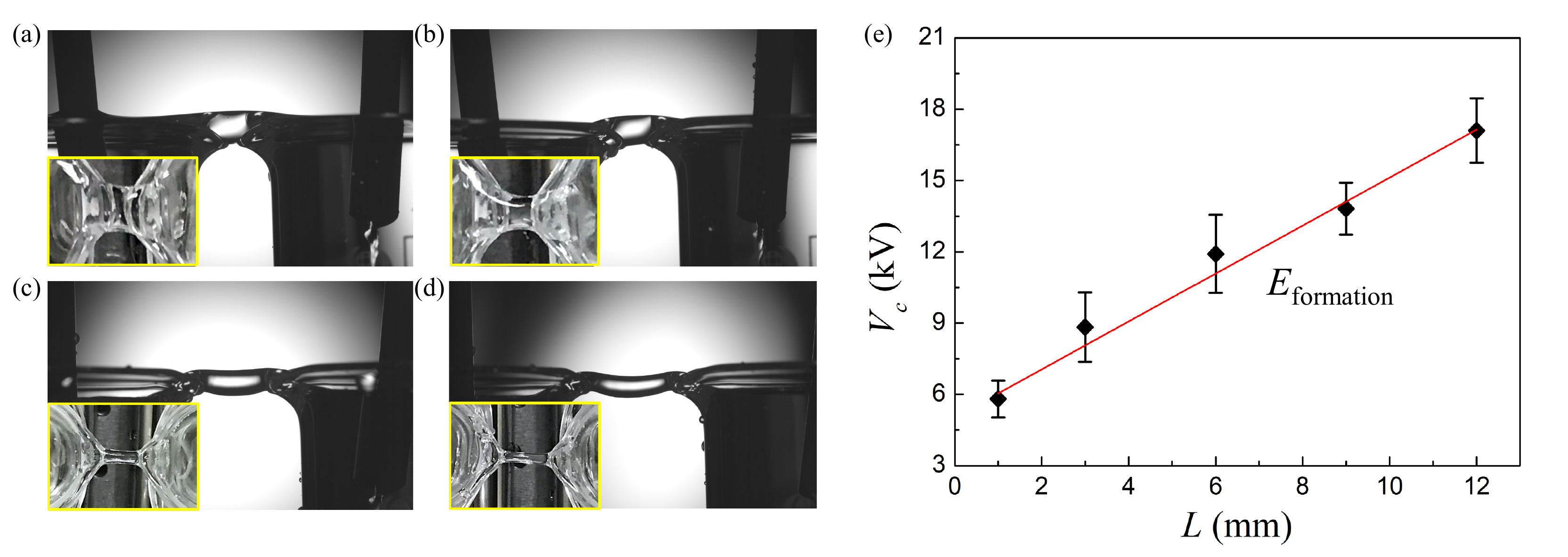}%
	\caption{Initial morphology under various $L$ and the associated $V_c$. (a-d) A side view of ALB morphology captured by the high-speed camera right after formation at the critical $V_c$ for various $L$, while the inset (the yellow rectangular) for a top-down view from the digital camera, $L=1, 3, 9, 12$ mm, and $V_c=6.0, 9.3, 14.7, 16.1$ kV for (a-d), respectively; (e)the critical voltage ($V_c$) increasing linearly with $L$, indicating a constant electric field ($E_\mathrm{formation} =1$ kV/mm) required for the formation (error bars for fifteen measurements).}
\label{fig:morphology}
\end{figure*}

By changing the distance $L$ between the two beakers, the morphology of ALB right after formation is captured (Fig.~\ref{fig:morphology}a-d). A larger distance $L$ requires a higher $V_c$, and the linear increase of $V_c$ with $L$ (Fig.~\ref{fig:morphology}e) indicates that a critical electric field is required to initiate the formation of ALB ($E_\mathrm{formation} \approx 1$ kV/mm).

Theoretically, in order to deform the water surface for the ALB formation, the Maxwell stress due to the electric field should be comparable with the surface tension,
\begin{equation}
 \epsilon\epsilon_0E^2\sim\gamma/\kappa^{-1},
\end{equation}
where $\gamma=73\times10^{-3}N/m$ is the surface tension of water, {the relevant length scale $\kappa^{-1} = 2.76$ mm  for the base diameter of the Taylor cone (Fig.~\ref{fig:formation}b at 10 ms)}, $\epsilon$ for the relative dielectric constant of the liquid, and $\epsilon_0$ for the vacuum permittivity \cite{Kimselective2011}. Then the estimated critical electric field ($E \sim \sqrt{\gamma/\epsilon\epsilon_0\kappa^{-1}}\sim1.64$ kV/mm) is fairly consistent with the experimental value of $E_\mathrm{formation}$.

For the typical electrocoalescence under an applied voltage \cite{BirdPRL2009,ChenPRL2013}, two droplets (with the diameter of hundreds of micrometers, and the distance about tens of micrometers) are deformed and connected. Here this liquid bridge is suspended between two macroscopic beakers (with the diameter of centimeters, and distance about millimeters), which can last or be stable for a while involving much more complicated physical phenomena.

\section{Mass transport via an axial flow of a waterfall}
\label{app:masstransport}
\begin{figure*}[t]
	\centering
	\includegraphics[width=\textwidth]{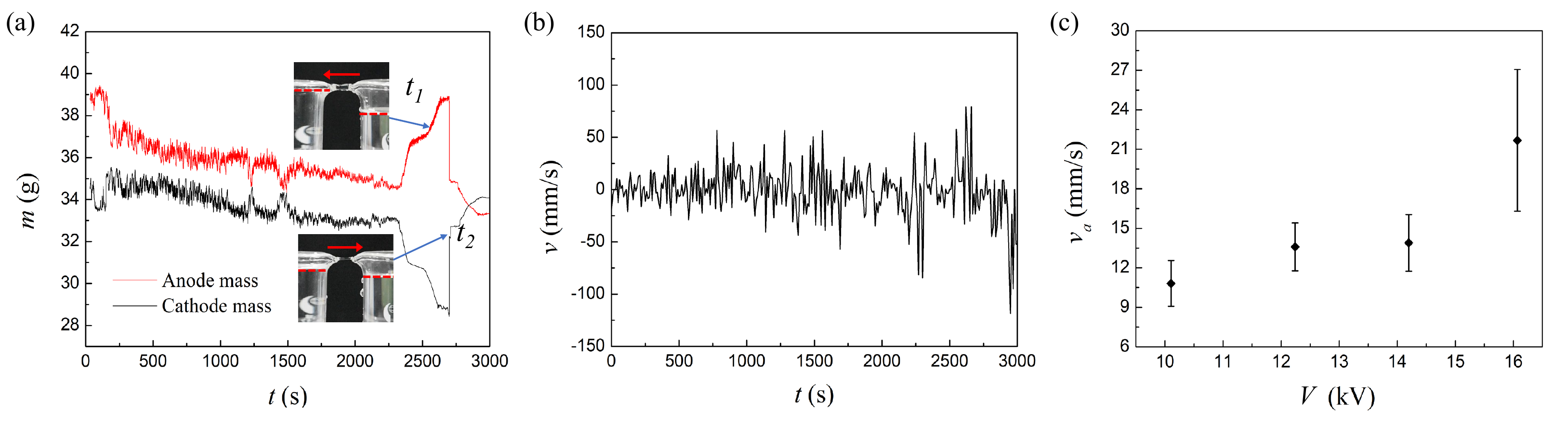}%
	\caption{{Mass transport between two beakers via a bidirectional axial flow. (a) The mass changing at the anode and cathode sides over time under 12 kV; (b) The flow velocity fluctuating around tens of millimeters per second; (c) The average velocity increasing with the applied voltage, on the order of tens of millimeters per second. All the experiments are conducted with the fixed distance $L$ = 6 mm and repeated four times ($L$ = 6 mm).}}
\label{fig:masschange}
\end{figure*}

In order to measure the mass change of water in the beakers in the anode and cathode sides, two beakers are placed on two electronic balances (S1700689 JC-1202B), which are connected to computers to collect the data.  The mass change with time on both sides is measured during the evolution of ALB. As shown in Fig.~\ref{fig:masschange}a, at time $t_1$, the liquid mass in the anode side is larger and is still increasing (see the red arrow in the upper inset), while at time $t_2$, there is a reversed flow of liquid from the anode side to the cathode side (see the red arrow in the lower inset), indicating a bi-directional flow during the evolution.

The flow velocity can be obtained from the mass change in Fig.~\ref{fig:masschange}a. The rate of mass change is
\begin{equation}
dm/dt =\frac{(m_1^2-m_2^2)/2-(m_1^1-m_2^1)/2}{t^2-t^1},
\end{equation}
where $m_i^j$ represents the mass at the time $t^j$ in the beaker $i$ (i=1,2 for the anode and cathode side respectively).
Then the flow velocity is
\begin{equation}
v=\frac{dm/dt}{\rho_wS^1},
\end{equation}
where $\rho_w$ is the density of the liquid and $S^1$ is the area of the liquid bridge at the time $t^1$. The flow velocity during the whole process fluctuates in the range of tens of millimeters per second as shown in Fig.~\ref{fig:masschange}b. The time-averaged velocity $\left | v \right |$ ($v_a$), in the order of tens of millimeters per second,  generally has a tendency to increase with the voltage (Fig.~\ref{fig:masschange}c).

\end{document}